\documentstyle[psfig,epsfig,aps,prb,twocolumn]{revtex}
\input{epsf}
\draft
\begin{document}
\title{AC-driven quantum spins:
resonant enhancement of
transverse DC magnetization}
\author{S. Flach, A. E. Miroshnichenko and A. A. Ovchinnikov}
\address{Max-Planck-Institut f\"ur Physik komplexer Systeme, N\"othnitzer
Strasse 38, D-01187 Dresden, Germany}
\date{\today}
\wideabs{
\maketitle
\begin{abstract}
We consider $s=1/2$ spins in the presence of a constant magnetic field 
in $z$-direction and an AC magnetic field in the $x-z$ plane.
A nonzero DC magnetization component in $y$ direction is a result
of broken symmetries. A pairwise interaction between two spins
is shown to resonantly increase the induced magnetization
by one order of magnitude. We discuss the mechanism 
of this enhancement, which is due to additional avoided crossings
in the level structure of the system.
\end{abstract}
}
\section{Introduction}

Magnetic resonance effects have been long studied and used for 
various spectroscopic methods \cite{fb56,nb65,pnbdc91}. Time-dependent external 
fields being an integral part of the physical setups, become also
very interesting in their action when entering the field of 
spintronics and processing of quantum information \cite{dgc00}.
While the linear response of a system (probe) to external fields
may be typically understood in a rather complete way, nonlinear
and perhaps nonadiabatic response effects are usually much harder
to be analyzed. Some studies have been devoted to the effect
of second-harmonic generation. Much less is known about 
{\sl zeroth-harmonic generation}, i.e. about the induction
of static fields by pure ac fields. The understanding of such processes
is very important in any application involving ac fields, as
additional induced dc fields may strongly influence the desired
effects.

The principal possibility of the zeroth-harmonic generation can be
studied using symmetry considerations 
\cite{sfoyyzl00},\cite{oysfyzaao01}. 
Analytical approaches which may also estimate the magnitude
of the induced fields are typically restricted to various perturbation
techniques with a suitable small parameter involved. 
Two of us have recently applied these methods to the simplest
cases of a spin $s=1/2$ coupled to an environment (heat bath)
in the presence of an external magnetic dc field (in $z$-direction)
\cite{sfaao01}.
While an additional ac field applied in $x$-direction leads to
the standard spin resonance setup
used e.g. for nuclear spins, a {\sl tilt} of the ac field direction
in the $x-z$-plane leads to a broken symmetry and to a generation
of a nonzero magnetic moment in $y$-direction, i.e. perpendicular
to the applied fields. Similar results have been also 
obtained for a spin $s=1$ where the external dc field in $z$-direction
is replaced by a magnetic ion anisotropy in $z$-direction \cite{sfaao01}.

This work is devoted to the analysis of spin-spin interaction
and its action on the above described effects. The first goal is
to study the generalization of the symmetry considerations in
the presence of an interaction. The second goal is to demonstrate
that the interaction is capable of increasing the induced field strength
by an order of magnitude, and to give an explanation for this result.

The paper is organized as follows. In the next section we introduce
the density matrix approach. Section III is devoted to a reconsideration
of the single spin problem. Section IV contains the central results on
two interacting spins. Section V concludes the paper.

\section{The density matrix approach}

In this section we introduce the methods we are using.
Let us assume that a single spin or a system of interacting spins
is represented by a time-dependent Hamiltonian $H(t)=H_0 + H_1(t)$
which includes both applied dc and ac magnetic fields.
Here the static part is completely included in $H_0$ and the
action of the ac fields is described by 
a time-periodic term $H_1(t)=H_1(t+T)$ with period $T$.
In order to describe the time-dependent statistical evolution
of such a system we use the quantum Liouville equation for
the density matrix $\rho(t)$
\begin{eqnarray}
\frac{\partial\rho}{\partial t}=i[H,\rho]-\nu(\rho-\rho_{e})
\label{2-1}
\end{eqnarray}
where $[A,B]=AB-BA$. 
While the specifics of the internal dynamics of the system are described
by the first term on the rhs of (\ref{2-1}), the second term accounts
for the interaction of our spin system with an environment.
We will handle this interaction in a rather qualitative way.
The principal request this coupling should fulfill is that its
action should preserve all symmetries of $H(t)$ except time reversal
symmetry. A second requirement is that the solution of (\ref{2-1})
in the absence of ac fields $H_1(t)=0$ should be the Gibbs distribution
\begin{eqnarray}
\rho_{\beta}(H_0)=\frac{1}{Z_0}e^{-\beta H_0}\;,\;Z_0=Tr(e^{-\beta H_0})\;\;.
\label{2-2}
\end{eqnarray}
Finally we request that all solutions of (\ref{2-1}) should approach
a unique time-dependent state.

The choice $\rho_e=\rho_{\beta}(H_0)$ satisfies the above criteria.
As shown in \cite{sfaao01} for $s=1/2$ it leads to equations practically identical
with the well known Bloch equations.
However e.g. for adiabatically slow ac fields a much better choice
would be $\rho_e \equiv \rho_{\beta}(H(t)) =  
\frac{1}{Z}e^{-\beta H(t)}$. We will discuss
the subtle differences below. In general however we may state here
that any of the two choices fulfills the above criteria.
Note that the third criterium (unique asymptotic behaviour) is
guaranteed, as a linear equation like (\ref{2-1}) exhibits one and only
one attractor, and the whole phase space serves as a region of attraction
for this single attractor.
Note that due to $Tr\rho_{\beta}=1$ any choice with $Tr\rho(0)=1$
implies $Tr\rho(t)=1$ for all $t$.

The value $\bar{A}(t)$ of an observable corresponding to the operator $A$ is
defined by
\begin{eqnarray}\label{observe}
\bar{A}(t)=Tr(A\rho(t))
\end{eqnarray}
and its time average is abbrevated by
\begin{eqnarray}
\tilde{A}=\frac{1}{T}\int_0^T\bar{A}(t)\,dt.
\end{eqnarray}

\section{Revisiting a single spin $s=\frac{1}{2}$}

We consider one spin $s=1/2$ in the presence 
of a constant magnetic field in $z$-direction and
a  time-periodic magnetic field with angle $\phi$ to the $z$-axis 
in $z-x$-plane. The Hamiltonian is given by
\begin{eqnarray}
H(t)=h_0S_z+h(t)(\alpha S_x+\gamma S_z),
\end{eqnarray}
where $\alpha=\sin\phi$ and $\gamma=\cos\phi$. 
We assume that the field $h(t)=h(t+T)$
has zero mean $\int_0^Th(t)\,dt=0$.

The spin component operators are given by the
Pauli matrices: $S_{x,y,z}=\frac{1}{2}\sigma_{x,y,z}$. The density matrix
$\rho=R+iI$ can be decomposed into a real symmetric $2\times 2$ matrix
$R$ and a real antisymmetric $2\times 2$ matrix $I$
and has three independent variables. Using the relation (\ref{observe}), we
rewrite the system (\ref{2-1}) in terms of $\bar{S}_{x,y,z}$
\begin{eqnarray}\label{sys1}
\dot{\bar{S}}_x&=&(h_0+\gamma h(t))\bar{S}_y-\nu\bar{S}_x\nonumber\\
\dot{\bar{S}}_y&=&\alpha h(t)\bar{S}_z-(h_0+\gamma
h(t))\bar{S}_x-\nu\bar{S}_y\\
\dot{\bar{S}}_z&=&-\alpha h(t)\bar{S}_y-\nu(\bar{S}_z+C)\nonumber
\end{eqnarray}
where $C=\frac{1}{2}\tanh(\beta h_0/2)$. 
The structure of these equations is equivalent to the wellknown 
Bloch equations provided the relaxation times there would be chosen
to be identical \cite{nb65}. Note that the symmetries of the equations (\ref{sys1})
are identical with the ones of the corresponding Bloch equations.

Because of the presence of a time-periodic field, for
times $t\gg1/\nu$ all components $\bar{S}_{x,y,z}$ will be also time periodic
with period $T$ or frequency $\omega=2\pi /T$. We can expand the solution of
system (\ref{sys1}) in a Fourier series
\begin{eqnarray}
\bar{S}_{x,y,z}=A_{0\;x,y,z}+\sum\limits_{n\not=0}e^{i\omega nt}A_{n\;x,y,z}
\;\;.
\end{eqnarray}
After averaging over the period $T$, we have 
\begin{eqnarray}
\tilde{S}_{x,y,z}=A_{0\;x,y,z}\;\;.
\end{eqnarray}

The symmetry analysis of the equations (\ref{sys1})
yields the following results \cite{sfaao01} (we considered only operations
which conserve $Tr \rho $):
\\
1) $\nu=0$, $\gamma=0\;,\;h(-t)=-h(t)$: 
\begin{eqnarray}\label{sym2}
\bar{S}_x\rightarrow-\bar{S}_x\;,\;\bar{S}_y\rightarrow\bar{S}_y\;,\;
\bar{S}_z\rightarrow\bar{S}_z\;,\;t\rightarrow -t.
\end{eqnarray}
2) $\nu=0$, for all $\gamma\;,\;h(-t)=h(t)$:
\begin{eqnarray}\label{sym3}
\bar{S}_x\rightarrow\bar{S}_x\;,\;\bar{S}_y\rightarrow-\bar{S}_y\;,\;
\bar{S}_z\rightarrow\bar{S}_z\;,\;t\rightarrow -t.
\end{eqnarray}
3) $\nu\not=0$, $\gamma=0,\;\;h(t+T/2)=-h(t)$:
\begin{eqnarray}\label{sym1}
\bar{S}_x\rightarrow-\bar{S}_x\;,\;\bar{S}_y\rightarrow-\bar{S}_y\;,\;
\bar{S}_z\rightarrow\bar{S}_z\;,\;t\rightarrow t+\frac{T}{2}\;.
\end{eqnarray}
If any of the above symmetries hold in a relevant parameter case,
the corresponding components of $\tilde{S}$ will exactly vanish.
E.g. for case 1) $\tilde{S}_x=0$, for case 2) $\tilde{S}_y=0$
and for case 3) $\tilde{S}_x = \tilde{S}_y=0$.
The key idea now is that a proper choice of the parameters (including 
the time dependence of $h(t)$) may lead to a violation of all symmetries.
As a consequence we expect 
nonzero
values of the corresponding $A_{0\;x,y}$.
This means that
even if we do not apply an external magnetic field in $y$-direction, 
we can expect a
nonzero $\tilde{S}_y$ component, which will be a function of the
ac field frequency $\omega$. In order to quantify the zeroth harmonic
generation we use the following definition of the strength of the effect
$I_{NL}$:
\begin{eqnarray}
I_{NL}=\frac{\max\limits_{\omega}|\tilde{S}_y|}{\max\limits_{
\omega}|\tilde{S}_z|}\cdot100\%
\label{inl}
\end{eqnarray}
Let us emphasize here that the denominator in (\ref{inl}) is of
nonresonant character, i.e. its value is determined by the maximum
values of $\tilde{S}_z$ which is obtained for
frequency values far from the magnetic resonance and practically coincides
with the equilibrium value for zero ac fields. In contrast the enumerator
gets its maximum (if nonzero et al) inside some resonance window in 
$\omega$.

In \cite{sfaao01} a perturbation theory was developed for strong dissipation
$1/\nu \ll 1$. Here we will account for the effect assuming a small
amplitude of the ac field
$h(t)=\epsilon\cos(\omega t)$, $\epsilon \ll 1$ and $\epsilon < \nu$:
\begin{eqnarray}
A_{0\;x}&=&-\epsilon^2\frac{\alpha\gamma C(\nu^2+\omega^2-h_0^2)}{2\Delta}+o(\epsilon^3)\label{ax}\\
A_{0\;y}&=&\epsilon^2\frac{\alpha\gamma \nu h_0C}{\Delta}+o(\epsilon^3)\label{ay}\\
A_{0\;z}&=&-C+\epsilon^2\frac{\alpha^2C(\nu^2+\omega^2+h_0^2)}{2\Delta}+o(\epsilon^3)\label{az}
\end{eqnarray}
where $\Delta=(h_0^2-\omega^2)^2+\nu^2(\nu^2+2h_0^2+2\omega^2)$.
The dependence of $A_{0\;x,y,z}$ on $\omega$ shows the expected 
resonant character.
A nonzero component of $\tilde{S}_y$, being a nonlinear response result,
appears in second
order of the ac field amplitude $\epsilon$.

For the case of the magnetic resonance $\omega=h_0$ we find
\begin{eqnarray}\label{w=h0-1}
A_{0\;x}&\approx&-\epsilon^2\frac{\alpha\gamma C}{2h_0^2(4+\xi^2)}\\
\label{w=h0-2}
A_{0\;y}&\approx&\epsilon^2\frac{\alpha\gamma C}{\xi h_0^2(4+\xi^2)}\
\end{eqnarray}
where $\xi=\frac{\nu}{h_0}$.

\begin{figure}[htb]
\vspace{20pt}
\centerline{\psfig{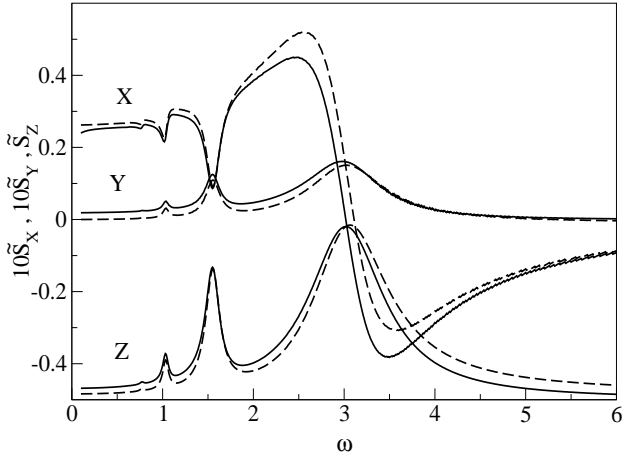}}
\vspace{2pt}
\caption{$\tilde{S}_{x,y,z}$ as function of $\omega$. Chosen parameters are:
$h_0=3.0$, $\phi=\frac{\pi}{4}$, $\beta=10$, $\nu=0.1$, 
$h(t)=\sqrt{2} \cos(\omega t)$. 
Solid lines: $\rho_e = \rho_{\beta}(H_0)$. Dashed lines: 
$\rho_e = \rho_{\beta}(H(t))$  }
\label{fig0}
\end{figure}

To test the possible magnitude of the effect, we compute the induced
moments numerically following \cite{sfaao01}. The results are the solid lines
in Fig.\ref{fig0}. While the resonance character is clearly observable, 
the strength $I_{NL}\approx3.0\%$ is rather small. 
%

In addition, one
can see that in the adiabatic limit
$\omega \rightarrow 0$ $\tilde{S}_y$ keeps a small but nonzero value.
This is in contrast to the expectation that for very slowly varying ac fields
the spin  system should momentarily adjust to the actual field value,
which implies a zero average $y$-component of the induced moment.
As already discussed above, a much better choice in this limit would
be $\rho_e = \rho_{\beta}(H(t))$.
A computation with this choice is shown also in Fig.1 (dashed lines).
We find that while the overall difference 
in the results
for the two choices for $\rho_e$
are small, the adiabatic limit is correctly reproduced by the second one
as expected.
%
\begin{figure}[htb]
\vspace{20pt}
\centerline{\psfig{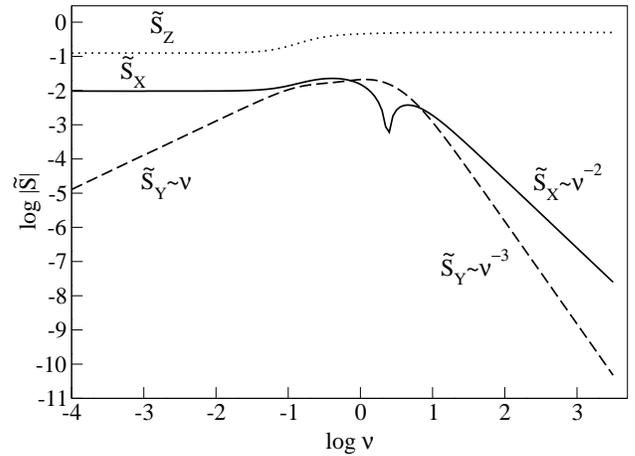}}
\vspace{2pt}
\caption{$\log \tilde{S}_{x,y,z}$ as function of $\log \nu$ for
$\omega=1.5$, other parameters as in Fig.\ref{fig0}. 
$\tilde{S}_x$: solid line, $\tilde{S}_y$ - dashed line, 
$\tilde{S}_z$ - dotted line.}
\label{fignu}
\end{figure}
It is also instructive to study the dependence of the values of
$\tilde{S}_{x,y,z}$ as a function of the dissipation $\nu$.
In Fig.\ref{fignu} we plot this dependence for the subresonance case
$\omega = 1.5$ from Fig.\ref{fig0}. We observe that $\tilde{S}_z$ shows
a kink around $\nu = 0.1$. The $\tilde{S}_x$ curve provides with the
a $\sim 1/\nu^2$ decay for large $\nu$, as expected from perturbation
theory. Except for a small dip around $\nu=2$ it then saturates
at a nonzero value for small $\nu$ values. This is in full agreement
with the above symmetry statements. Finally the $\tilde{S}_y$ curve
shows a $\sim 1/\nu^3$ decay for large $\nu$, again as expected
from perturbation theory. Notably $\tilde{S}_y$ tends to zero
also for small dissipation values, in full accord with the above
symmetry statements.

\section{Two interacting spins}

In order to enhance the value of $I_{NL}$ we may either consider
larger spins or effects of interactions between spins. In \cite{sfaao01}
the case of $s=1$ has been considered, with the result that $I_{NL}$
is not significantly changed as compared to $s=1/2$. Consequently
we turn in this section to the consideration of spin interactions.
In the following we will study a system of two interacting spins $s=1/2$,
which could be realized through the interactions e.g. in a molecule.
As we are interested in the qualitative effects such an interaction might
induce, we choose the mathematically simple version of exchange interaction.
We think that e.g. dipole-dipole interactions as realized between nuclear
spins will not significantly alter the picture.

Our starting Hamiltonian is given by
\begin{eqnarray} \label{s1s2}
H&=&h_0(S_z^1+S_z^2)+h(t)(\alpha(S_x^1+S_x^2)+\gamma(S_z^1+S_z^2))-\nonumber\\
&-&J_xS_x^1S_x^2-J_yS_y^1S_y^2-J_zS_z^1S_z^2
\end{eqnarray}
with $h(t+T)=h(t)$, $\alpha=\sin(\phi)$ and $\gamma=\cos(\phi)$.
For two spins $s=1/2$ the above operators $S_{x,z}^{1,2}$ take the form
$S_{x,z}^1=\frac{1}{2}\sigma_{x,z}\bigotimes\mathbf{1}$, 
$S_{x,z}^2=\mathbf{1}\mathnormal\bigotimes\frac{1}{2}\sigma_{x,z}$
where $\sigma_{x,z}$ are Pauli matrices, $\mathbf{1}$ denotes
the $2\times 2$ unity matrix
and $\bigotimes$ stands for the tensorial product of operators (matrices).

In this case the matrix $\rho=R+iI$ contains 15 independent variables.

We define the total components of spins as
$S_{x,y,z}=\frac{1}{2}(S_{x,y,z}^1+S_{x,y,z}^2)$. 
With the help of (\ref{observe}) it follows
\begin{eqnarray}
\bar{S}_x&=&\frac{1}{2}(R_{12}+R_{13}+R_{24}+R_{34})\nonumber\\
\bar{S}_y&=&\frac{1}{2}(I_{12}+I_{13}+I_{24}+I_{34})\\
\bar{S}_z&=&\frac{1}{2}(R_{11}-R_{44})\nonumber
\end{eqnarray}

The symmetry analysis of this system is identical
to the case of one spin $s=1/2$ 
(\ref{sym1}-\ref{sym3}). The existence of additional parameters (exchange
coefficients) does not reduce the symmetries nor add some.
The only important additional point is that changes of signs of 
the expectation values of a given spin component operator have to be
done simultaneously for both spins once the interaction is nonzero.

\subsection{Isotropic exchange}

Let us first consider the case of isotropic exchange
$J_x=J_y=J_z\equiv J$. 

First we note that with the help of a unitary transformation
we can transform the Hamiltonian (\ref{s1s2}) into a
triplet-singlet representation. The singlet state (total spin $s=0$)
decouples from the three triplet states (total spin $s=1$).
The time-dependent part of (\ref{s1s2}) is contained in the triplet
part.
For a given value of the ac field the eigenvalues of the singlet
state $E_0$ and the three triplet states $E_{1,2,3}$ are given by
\begin{eqnarray}
E_0&=&\frac{3}{4}J\nonumber\\
E_1&=&-\frac{1}{4}J+\sqrt{a^2+4c^2},\nonumber\\
E_2&=&-\frac{1}{4}J,\\
E_3&=&-\frac{1}{4}J-\sqrt{a^2+4c^2} \label{levels}
\end{eqnarray}
where $a=h_0+\gamma\cos(\omega t)$ and
$c=\frac{1}{2}\alpha\cos(\omega t)$. 

On one side the singlet state is not interacting with the triplet states
and all spin component expectation values in this singlet state are 
exactly zero. On the other side the singlet state is consuming statistical
weight.
Thus the dynamics of the system of two spins one-half 
interacting via isotropic exchange and coupled to a heat bath
can be reduced to the dynamics
of a single spin $s=1$  (triplet states) with a reduced statistical weight.
For the case of $\rho_e=\rho_{\beta}(H_0)$
this reduced statistical weight is given by  
\begin{equation}
C_1=\frac{Tr(e^{-\beta H_0^{\prime}})}{Tr(e^{-\beta H_0})}
\end{equation}
where $H_0^{\prime}$ is obtained from
$H_0$ by excluding the singlet state.
Using
$Tr(e^{-\beta H_0^{\prime}})=2e^{\frac{\beta J}{4}}\cosh(\beta h_0)
+e^{\frac{\beta J}{4}}$
and $Tr(e^{-\beta H_0})=Tr(e^{-\beta H_0^{\prime}})+e^{-\frac{3\beta J}{4}}$
we find
\begin{eqnarray} \label{c1}
C_1=\frac{2\cosh(\beta h_0)+1}{2\cosh(\beta h_0)+1+e^{-\beta J}}\;\;.
\end{eqnarray}
The coefficient (\ref{c1}) tends to $1$ for large ferromagnetic interaction
$J\rightarrow\infty$ and to $0$ for large anti-ferromagnetic interaction
$J\rightarrow-\infty$. 

As we know the value of $I_{NL}$ for $J=0$ (it is equivalent to
the case of a single spin $s=1/2$ discussed in the previous section)
we may obtain the values for $I_{NL}$ and all other numbers including
curves as in Fig.1 using (\ref{c1}). First we note that antiferromagnetic
interactions $J~<~0$ reduce the effect by making the singlet
state energetically favourable. This is reasonable as
for zero interaction both spins will on average point to some direction.
Antiferromagnetic interaction will favour antiparallel spin orientations
and thus reduce the induced moments. 
However ferromagnetic interaction, making the singlet state energetically
unfavourable, will enhance the effect. An upper bound for the
maximum enhancement can be easily obtained from (\ref{c1}).
In the most favourable case the statistical weight of the triplet states
will increase up to $4/3$ of the $J=0$ case. Thus the case of
isotropic ferromagnetic exchange, while leading to an enhancement of
the induced moments, does not change the results drastically as compared
to zero interaction. We will now turn to anisotropic exchange interaction
and see that the situation is going to change dramatically.

\subsection{Anisotropic exchange}

To provide with a systematics in the search for exchange values which
may significantly enhance the induced magnetic moments we first mention
that as for isotropic exchange, also the general case allows for a 
unitary transformation which separates a singlet state (total spin $s=0$)
from three triplet states (total spin $s=1$). The singlet state
will become unimportant if its consumed statistical weight is small.
This can be typically obtained by choosing ferromagnetic exchange
interactions. 
Second we recall that for isotropic exchange 
the dependence of the triplet eigenenergies
for a given value of the ac field $h$ (frozen time, or adiabatic limit)
as a function of this ac field value $h$ will show up with a pair of states
performing an avoided crossing at $h=0$ and a third state being independent
of $h$. This follows immediately from (\ref{levels}).
The minimum distance between the two states $E_1$ and $E_3$ is given
by the Zeeman splitting $2h_0$, as in the case of a single spin $s=1/2$
(up to the factor 2). Also we note that the ac field term induces interactions
and thus transitions between these two states, leading to the 
standard magnetic resonance. 

For anisotropic exchange the situation changes. The third level $E_2$,
which was not participating in the resonance dynamics for isotropic
exchange, now gets involved in the dynamics due to 
additional nonzero matrix elements for anisotropic exchange.
As a consequence we may expect resonant response at different
frequencies (energies) corresponding to the different level spacings
$E_1-E_3$ and $E_2-E_3$. Especially important is, that we may consider
e.g. a resonant response for transitions from the lowest level to the highest
one of the triplet states, 
while having at the same time a narrowly avoided crossing
between the lowest state and the intermediate one. The idea behind this
scenario is that i) small energy differences in the vicinity of the
avoided crossing will lead to small denominators in the equation for
the density matrix and ii) an additional avoided crossing provides
with a strong nonlinear dependence of the wavefunction of the lowest
triplet state upon variation of $h$. So our strategy is to find
exchange values which realize such a situation. Then we choose
amplitudes of the ac field which will cover the avoided crossing,
and temperatures small enough such that even in the point of the
avoided crossing the statistical weight of the intermediate
state is small as compared to the triplet ground state.
 
\begin{figure}[htb]
\vspace{20pt}
\centerline{\psfig{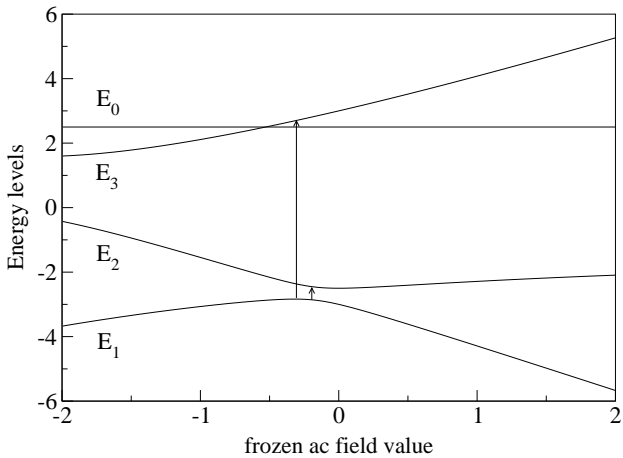}}
\vspace{2pt}
\caption{Energy level dependence on frozen ac field value
(parameters see text). Arrows indicate two possible resonant
excitations. }
\label{fig2}
\end{figure}

The example we will demonstrate corresponds to a ferromagnetic easy plane
exchange interaction 
$J_x=J_y=5.0$ and $J_z=0$. 
The dependence of the levels on the amplitude of the frozen ac field
is shown in Fig.\ref{fig2}. 
%
The minimum distance between the two lowest states in the avoided crossing
region is $0.5$. 
We now expect a possible strong enhancement 
of the induced moments in the frequency range of the resonance
indicated by the arrow $\omega \sim 5.8$ which should be
most effective if the temperature is smaller than the avoided crossing
splitting, i.e. $\beta~>~2$.
In addition we expect to see another resonance at frequencies
corresponding to the smallest splitting between the two lowest lying
states, i.e. around $\omega = 0.5$. 
%
\begin{figure}[htb]
\vspace{20pt}
\centerline{\psfig{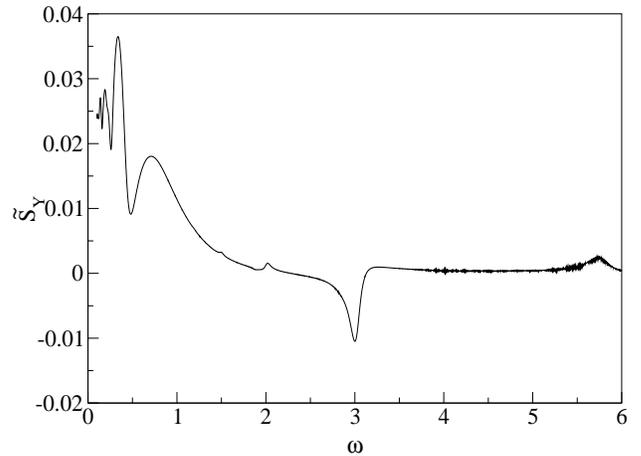}}
\vspace{2pt}
\caption{$\tilde{S}_{y}$ as function of $\omega$.
Chosen parameters are:
$h_0=3.0$, $\phi=\frac{\pi}{4}$, $\beta=10.0$, $\nu=0.1$, $J_x=J_y=5.0$ and
$J_z=0.0$. In this case $h(t)=\sqrt{2} \cos(\omega t)$.}
\label{fig3}
\end{figure}
In Fig.\ref{fig3} the results of numerical simulation for $\tilde{S}_y$
are shown
for $\beta=10$ and $\nu=0.1$. We do not observe a strong 
response at the expected frequency value 5.8, but we do observe in addition
to the standard response at $\omega=h_0=3$ a response at $\omega \approx 0.5$
which corresponds to the resonant excitation in the region of the narrow
avoided crossing in Fig.\ref{fig2}. The value of $I_{NL}$ is already pretty large
here, of the order of 7\%. 

\begin{figure}[htb]
\vspace{20pt}
\centerline{\psfig{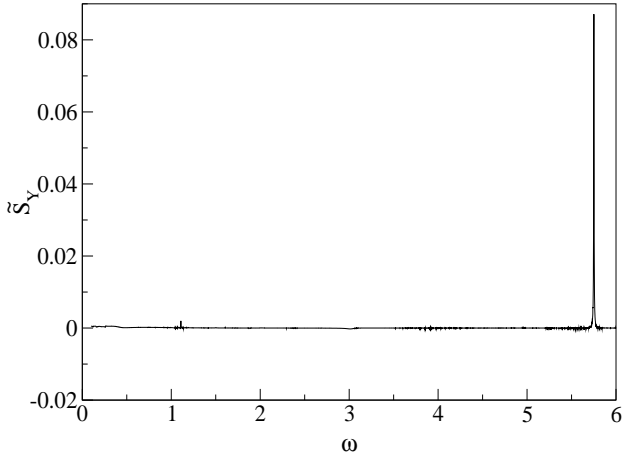}}
\vspace{2pt}
\caption{$\tilde{S}_{y}$ as function of $\omega$. 
Parameters as in Fig.\ref{fig3} except 
$\nu=10^{-3}$}
\label{fig4}
\end{figure}

\begin{figure}[htb]
\vspace{20pt}
\centerline{\psfig{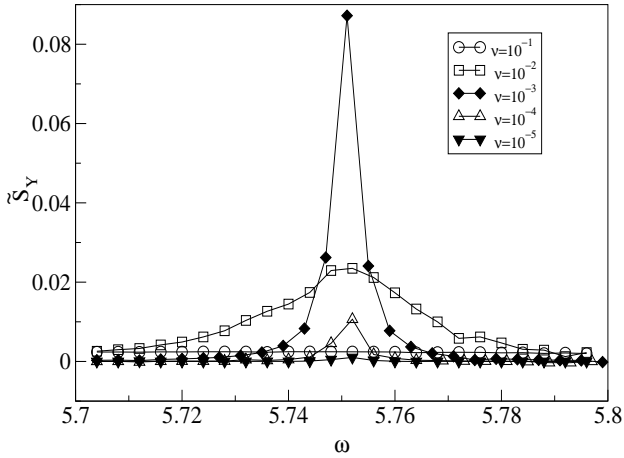}}
\vspace{2pt}
\caption{
$\tilde{S}_y$ versus $\omega$ in the strong resonance region $\omega \approx 
5.8$ for different values of $\nu$. Other parameters as in Fig.\ref{fig3}.}
\label{fig5}
\end{figure}

\begin{figure}[htb]
\vspace{20pt}
\centerline{\psfig{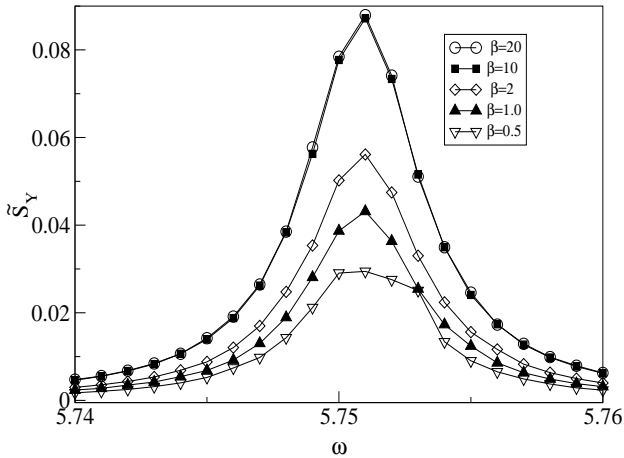}}
\vspace{2pt}
\caption{
$\tilde{S}_y$ versus $\omega$ in the strong resonance region $\omega \approx 
5.8$ for different values of $\beta$. Other parameters as in Fig.\ref{fig3}
except $\nu=10^{-3}$.}
\label{fig6}
\end{figure}

In order to enhance the response at $\omega \approx 5.8$, 
we decrease the value for the dissipation constant
by two orders of magnitude to $\nu=10^{-3}$. The resulting curve
is shown in Fig.\ref{fig4}. We clearly find a strong enhancement of
the resonance at $\omega \approx 5.8$. In this case the value
of $I_{NL} \approx 18\%$, which is the largest value we achieved
to find.
To characterize the dependence of this strong resonance on the parameters
$\beta$ and $\nu$ we plot different curves for additional values 
 of  $\nu$ (Fig.~\ref{fig5}) and  $\beta$ (Fig.~\ref{fig6}).
The dependence on the inverse temperature shows the expected saturation
at low temperatures. The dependence on $\nu$ shows a surprising
non-monotonic behaviour. While it would be tempting to discuss it
in terms of so-called stochastic resonance, we think that a much simpler
explanation is due to the fact that in the limit of zero dissipation
time reversal symmetry (\ref{sym3}) will be restored. Consequently
in the limit $\nu \rightarrow 0$ the values of the $y$-component
of the induced dc magnetic moment will vanish for all frequencies.
%
%
%

\section{Conclusion}

In conclusion we would like to stress once again that strong AC magnetic
fields may generate zeroth harmonics
in a spin system.
This effect has resonant behaviour as a function of the AC field 
frequency and could be a new tool for the spectroscopic
study of magnetic systems.

We have shown that the induction of a dc magnetic moment perpendicular
to applied magnetic fields can be significantly enhanced for
systems of interacting spin pairs as compared to single spins.
This enhancement of the nonadiabatic nonlinear response effect
is due to the presence of additional energy levels in a spin pair system,
which allows to obtain additional avoided crossings between levels
upon applying external fields. These avoided crossings strongly
contribute to the nonlinear response through the dramatic change of
wavefunctions as one sweeps through such a crossing.
The presence of avoided crossings is typically associated with chaos
and nonintegrability in corresponding classical systems. 
The case of isotropic exchange reduces the system to a single spin
one and consequently does not lead to a strong enhancement
of the induced moments. 
 
The particular case of two interacting spins could be realized in stable
biradicals, which provides us with the system of two spins located on the same
molecule.
Experimental possibility of observation of the above effect could be connected
with diluted paramagnetic media at very low temperature, because only at low
temperature the rate of dissipation of energy $\nu$ may become
very small,
thus, preventing the heating up of the system and narrowing the width of
resonance, which is very important for the observation.

Other candidates for such an observation are dielectric glasses based on
$SiO_x$. The low temperature properties of such glasses are
connected with the existence of systems of double-well tunnel centers. The
dynamics of these tunnel centers maps exactly onto the dynamics of 
spin systems. Thus, 
under the action of an AC electric field 
a permanent polarisability component could be induced.
Again low temperatures are a key requirement for the observation of this
effect.
\\
\\ 
Acknowledgements
\\
We thank M. V. Fistul for helpful discussions and a critical
reading of the manuscript.
This work was supported by the Deutsche Forschungsgemeinschaft
FL200/8-1.

\end{document}